\documentclass[%
 aip,
jcp,
 amsmath,amssymb,
reprint,
]{revtex4-1}

\usepackage{graphicx}
\usepackage{dcolumn}
\usepackage{bm}

\usepackage[utf8]{inputenc}
\usepackage[T1]{fontenc}
\usepackage{mathptmx}
\usepackage{etoolbox}
\usepackage{sidecap}
\usepackage{xcolor}

\makeatletter
\def\@email#1#2{%
 \endgroup
 \patchcmd{\titleblock@produce}
  {\frontmatter@RRAPformat}
  {\frontmatter@RRAPformat{\produce@RRAP{*#1\href{mailto:#2}{#2}}}\frontmatter@RRAPformat}
  {}{}
}%

\makeatother
\begin{document}
\newcommand{\para}[1]{{\color{red}{\tt #1}}}

\title[]{Random spin committee approach for smooth interatomic potentials}
\author{Vlad Cărare}

 \email{vc381@cam.ac.uk}
\affiliation{ 
Engineering Laboratory, University of Cambridge,\\ Trumpington St and JJ Thomson Ave, Cambridge, UK
}%
\author{Volker L. Deringer}%
 
\affiliation{%
Inorganic Chemistry Laboratory, Department of Chemistry,\\ University of Oxford, Oxford OX1 3QR, UK
}%

\author{Gábor Csányi}

\affiliation{ 
Engineering Laboratory, University of Cambridge,\\ Trumpington St and JJ Thomson Ave, Cambridge, UK
}%

\date{\today}

\begin{abstract}

Training interatomic potentials for spin-polarized systems continues to be a difficult task for the molecular modeling community. In this note, a proof-of-concept, random initial spin committee approach is proposed for obtaining the ground state of spin-polarized systems with a controllable degree of accuracy. The approach is tested on two toy models of elemental sulfur where the exact optimal spin configuration can be known. Machine-learning potentials are trained on the resulting data, and increasingly accurate fits with respect to the ground state are achieved, marking a step towards machine-learning force fields for general bulk spin-polarized systems.
\end{abstract}

\maketitle

Spin-polarization plays an important role in a variety of systems, such as elemental iron \cite{jana_searching_2023}, iron-containing compounds\cite{nejadsattari_spin_2021}, transition metal compounds more broadly\cite{reiher_theoretical_2009}, polymeric liquid sulfur and superheated liquid sulfur \cite{gardner_paramagnetic_1956,plasienka_structural_2015,munejiri_structural_2019}, but also in radicals as spin-probes for bio-macro-molecules\cite{jeschke_distance_2007}, to name a few examples. Building machine-learning (ML) force fields \cite{deringer_machine_2019,noe_machine_2020,unke_machine_2021,deringer_gaussian_2021,kocer_neural_2022} for such cases is not a straightforward task, as most atomistic ML architectures \cite{bartok_gaussian_2010,drautz_atomic_2019,batzner_advancing_2023,batatia_mace_2022} do not account for electronic spin degrees of freedom, although efforts have been made in this direction \cite{drautz_atomic_2020,novikov_magnetic_2022,shenoy_collinear-spin_2024}. For the former, simpler category of models, feeding atomic configurations of the same geometry and elements that are not in the same spin state can trivially lead to poor fits. Here we are concerned with tackling this issue from a data perspective, by proposing a simple protocol for finding the electronic spin ground state of an atomistic system. We note that this is expected to work best for cases where excited spin states are not widely populated. We work with two toy models of elemental sulfur, where the spin ground state can be exactly known. We show that we can fit smooth and accurate ML models by training on a best-of-many density functional theory (DFT) calculations where the initial atomic spins are randomly initialized and best-of is defined as taking the lowest energy DFT result for a given atomic geometry.

\begin{SCfigure}
\includegraphics[width=0.23\textwidth]{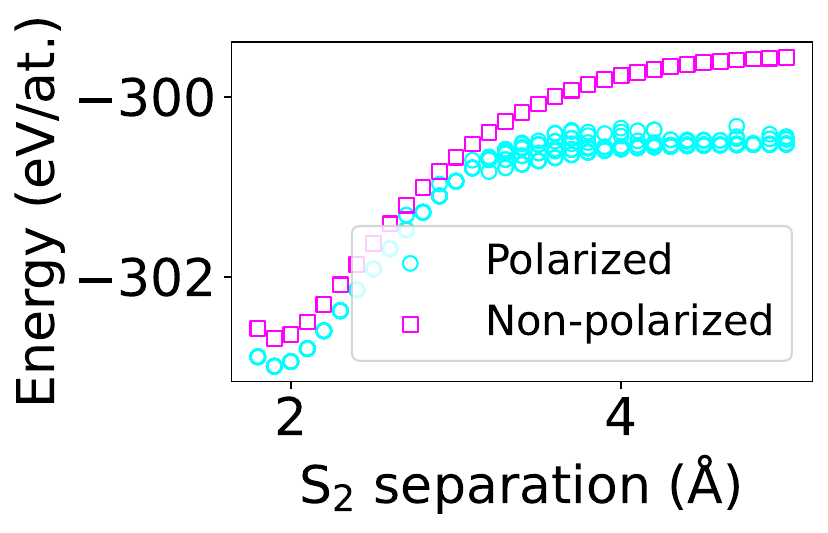}
\caption{\label{fig:S2} DFT energy profile along the S$_2$ dimer dissociation for non-spin-polarized (magenta) and spin-polarized with random initial spins (cyan) calculations. }
\vspace{-0.5cm}
\end{SCfigure}

\begin{figure}
\includegraphics[width=0.48\textwidth]{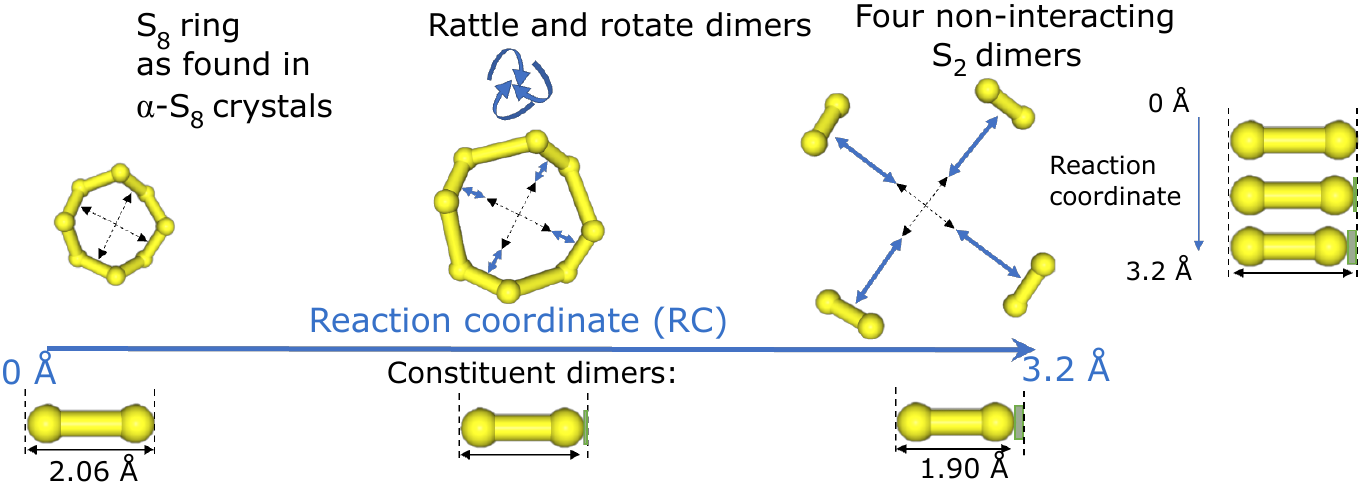}
\caption{\label{fig:S8dissociationSketch}Sketch of the toy model S$_{8}$ dissociation. The S$_6$ toy model follows a similar reaction, but ending in 3 dimers. We only rattle and rotate dimers in the rattled molecule dissociations (used for training and testing) and not in the rigid molecule dissociation (used solely for testing).}
\vspace{-0.5cm}
\end{figure}

\quad The first toy model that we study is that of a dissociation of an S$_{8}$ ring, which is in a singlet spin state in the ground state, to 4 S$_{2}$ dimers, which are in a triplet state, akin to an oxygen dimer \cite{steudel_elemental_II_2003} (we show in Fig. \ref{fig:S2} the energies of various spin states along the dimer dissociation). S$_{8}$ is the most common sulfur cluster found in nature \cite{steudel_elemental_2003}, in the solid, liquid and gas phases. Liquid sulfur undergoes temperature-induced polymerization \cite{steudel_elemental_2003}, where most S$_{8}$ rings begin to break and form chains or larger rings. The toy model discussed here is meant to mimic the first step in this process. The energetics of the S$_{8}$ to 4 S$_{2}$ equilibrium reaction were reviewed in chapter 3 of ref. \cite{steudel_elemental_2003}, although the exact reaction mechanism was not discussed. We obtained the experimental molecular crystal structure data for S$_{8}$, in its stable, ambient form from ref. \cite{george_lattice_2016}. We identified the coordinates of an S$_{8}$ ring and used ASE \cite{hjorth_larsen_atomic_2017} to create an isolated copy within a 24 \AA\ supercell. We then paired up the 8 atoms and moved them away from the center of the ring, in unison. We define the \textit{reaction coordinate} (RC) to be the amount by which the 4 dimers have moved away from their original positions. As they advance, we linearly decrease the \textit{reference} bond length from 2.06 \AA, the distance between atoms in the S$_{8}$ ring, to 1.90 \AA, the isolated dimer bond length, over a window of 1.90 \AA\ in reaction coordinate, which is the separation after which we found that the dimers stop influencing each other (the dimer's assigned magnetic moments became well defined integers under Mulliken analysis\cite{segall_population_1996} in CASTEP\cite{clark_first_2005}). We will call this the \textit{rigid molecule dissociation} of the S$_{8}$ ring and use it solely as a first test for the potentials. In order to create a training set (and a secondary testing set), we take this reaction and we rotate the dimer bonds and rattle the atoms, such that we have multiple data points at each value of the reaction coordinate. We call this the \textit{rattled molecule dissociation}. This was created so that the training set environments bear some similarity to the testing ones. 
The S$_{8}$ rattled molecule dissociation dataset is composed of 52 snapshots, where the reference bond length of each dimer follows a linear decrease from 2.06 to 1.90 \AA\ over 3.20 \AA\ in RC (although we stop collecting data at RC 2.70 \AA). Each dimer bond is further randomized by stretching or compressing it by a term from the uniform distribution bounded by $\pm$0.05 \AA\, and rotating it in a random direction by 1, 2, 3 and 4 degrees, such that at each reference reaction coordinate we have 4 structures. We keep every 4th structure for testing (in addition to the rigid molecule dissociation dataset), and use the rest for training. We chose GAP \cite{bartok_gaussian_2010} as the ML fitting tool to use here, although the conclusions are expected to be method-independent. We show in Fig. \ref{fig:S8dissociationSketch} a sketch of the implemented reaction.

The energies of the structures were evaluated under the DFT framework, in CASTEP 19.11 \cite{clark_first_2005}, with a 250 eV cutoff for the plane wave energy, 0.04 A$^{-1}$ $k$-mesh spacing, a 24 \AA\ cubic cell, PBE \cite{perdew_generalized_1996} functional and total spin fixed for 1 SCF cycle (meaning the initial magnetic moment value is relaxed after one self-consistent-field step). Spin-polarized calculations (in the collinear spin treatment) incur a doubling of the computational time, as the SCF cycles are ran separately for up and down occupancies. We find that setting the initial atomic spins on adjacent dimers to opposite values (we call this \textit{anti-parallel initialization}) allows to deterministically find a smooth, continuous energy curve for the S$_8$ reaction, which we hypothesize traces the ground state.

The second toy model we discuss is that of a  S$_{6}$ ring dissociation to 3 S$_2$ dimers. We obtained the S$_6$ trigonal crystal structure from Materials Project\cite{jain_commentary_2013} identifier \textit{mp-7} (which uses results from ref. \cite{steidel_redetermination_1978}) and isolated an S$_{6}$ ring from it. Then we implemented \textit{rigid} (1 structure per reaction coordinate) and \textit{rattled} (more structures per formal RC) molecule dissociation reactions similar  to the S$_{8}$ case above (cf. Fig. \ref{fig:S8dissociationSketch}), but with the reference bond length changing from 2.07 \AA\, for the ring, to 1.90 \AA\ for the isolated dimer, over a window of 1.20 \AA\ along the reaction coordinate. We chose S$_6$ for several reasons. Firstly, the S$_6$ ring is, along with S$_{8}$, one of the most stable sulfur molecules\cite{meyer_elemental_1976}. Secondly, given that the S$_6$ ring is, akin to S$_8$, also non-spin-polarized in the ground state, a dissociation into 3 dimers initialized in an anti-parallel spin arrangement results in a spin-polarized frustrated system at intermediate RCs. In this regime only, we find that a set-up of 2 anti-parallel dimers and one non-spin-polarized dimer traces the lowest energy curve.

Consequently, there exists no simple universal ground-state-finding strategy for setting the initial atomic spins that works deterministically for both the S$_{8}$ and S$_{6}$ dissociations, with the general case of a condensed system seemingly hopeless\cite{whitfield_NP-completeness_2014}. In response to this, we propose a scheme that makes use of the randomness of the SCF relaxation: instead of running one spin-polarized calculation per structure, we show that having a committee of $n$ calculations, each with different random initial atomic spins valued between $\pm\hbar/2$, and taking the lowest energy one post-relaxation (\textit{best-of-n}), is a reliable way to obtain the ground state. We call this the \textit{spin committee approach}. Note that the computational cost scales as \textit{n}. We point to the success of the Ab-Initio Random Structure Search methods pioneered by Pickard and Needs\cite{pickard_ab_2011} as an indication of why this may work. The authors argue empirically that in a general energy landscape, the lowest lying states are also part of the largest basins. This allows random initial guesses to find the optimal energy with accuracy. It is possible that this also holds in the space that the SCF procedure works in.

Naturally, the question follows as to how many spin committee members suffice to find the ground state. This comes down to a balance between the desired accuracy and the available resources. We show that in the case of the sulfur toy models mentioned here, the ML predictions converge after adding more than 5 committee members. However, it is plausible that for larger and different systems more members may be necessary. 

\begin{figure}
\includegraphics[width=0.48\textwidth]{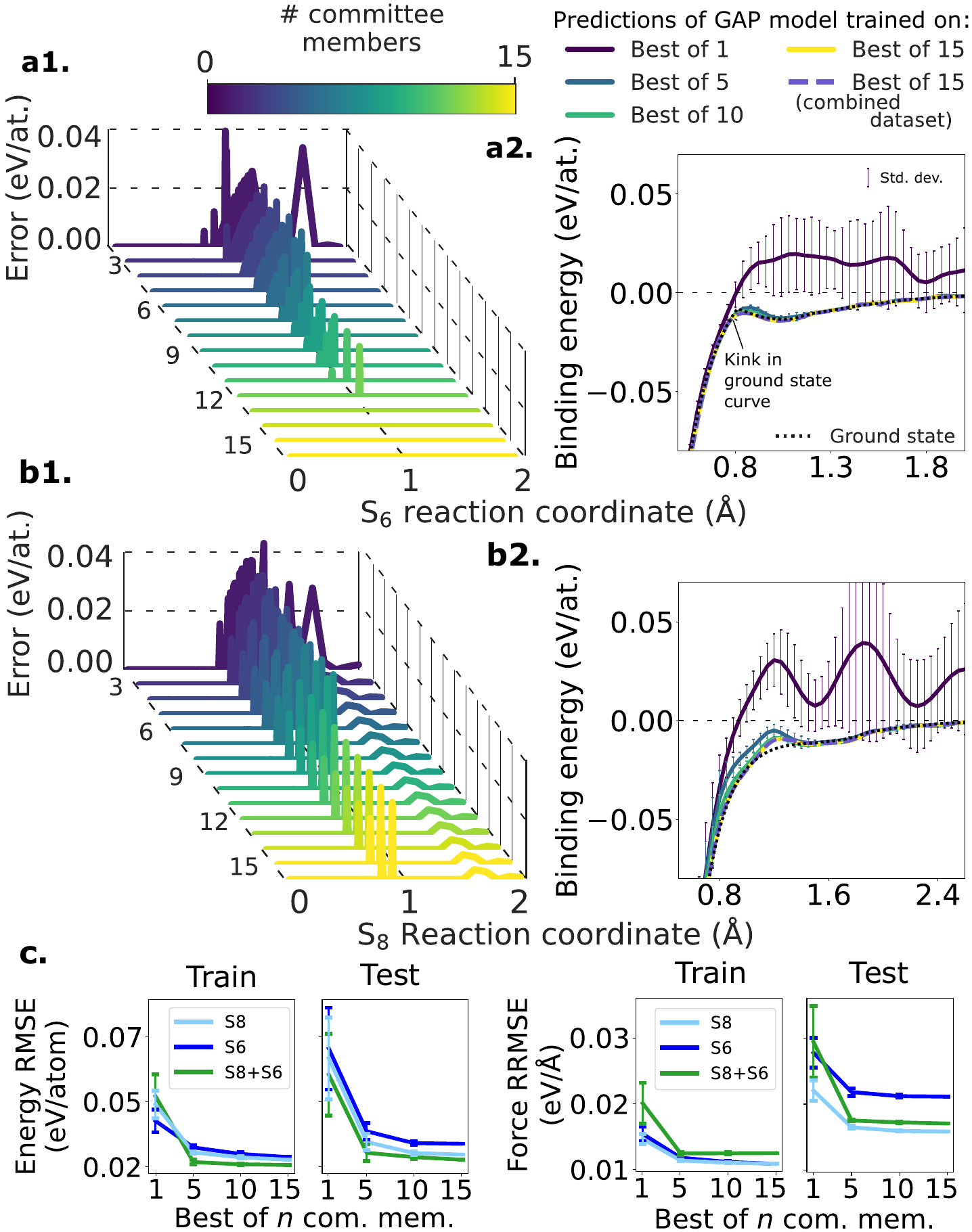}
\caption{\label{fig:S6andS8diss-spinCommitteeEvolution-Rigid.png} Reference and predicted results for the spin committee approach. \textbf{a1}: We plot, along the S$_{6}$ \textit{rigid} molecule dissociation (one structure per RC), the evolution of errors of the lowest DFT energy of increasingly larger spin committees with respect to the lowest DFT energy obtained from a meta-committee composed of the best-of-15 and of the anti-parallel states. \textbf{a2}: The predicted energy curves for the S$_{6}$ rigid molecule dissociation for GAP models trained on best-of-1, 5, 10 and 15 random spin committee members. The error bars are explained in the text. \textbf{b.}: Respective plots for the S$_{8}$ dissociation. \textbf{c}: Training and testing learning curves for 3 different \textit{rattled} molecule dissociation databases: S$_{6}$, S$_{8}$, and the combined S$_{6}$ S$_{8}$ set;  as a function of number of random spin committees per structure from which the lowest energy one was selected.}
\vspace{-0.5cm}
\end{figure}

In Fig. \ref{fig:S6andS8diss-spinCommitteeEvolution-Rigid.png} \textbf{a1}, we plot the evolution of DFT binding energy errors along the S$_{6}$ rigid (1 structure per RC) molecule dissociation as we increase the number of spin committee members we sample the best from. The equivalent plot for S$_8$ is shown in panel \textbf{b1}. The baseline binding energy in both cases is that which is the lowest between the best-of-15 and the anti-parallel initialization cases. For each rigid molecule structure we stop adding committee members before reaching a total of 15 if any calculation has reached the hypothesized ground state (this is not the case for the rattled molecule dissociations). From the plot, it is clear that the errors decrease with increasing number of spin committee members, and the ground state is recovered. 
Situations where the SCF cycles do not converge are frequent in the frustrated regime near 1 \AA\, where up to half of the computations can fail. This may be the reason for the best-of-15 S$_{8}$ curve (yellow line) not flattening, along with the larger number of atoms compared to S$_6$. One can additionally notice a non-smooth, kink feature in the S$_6$ ground state of panel \textbf{a2}. This is the sign of a crossing of electronic spin states, from a frustrated state where two dimers are anti-parallel, with the third one being non-spin-polarized, to a state were all 3 dimers are in a spin-polarized, triplet state.

We can test the random spin initialization scheme by comparing GAP fits to rattled S$_{6}$ and S$_{8}$ molecule dissociations composed of best-of-1 (committee members per structure), 5, 10 and 15.
To provide meaningful error bars on our result we create 15 different models for each of the best-of-1, 5 and 10 cases. These can be acquired by picking, for each structure, different committee combinations out of the total of 15 members. In the case of best-of-15, there is trivially only one way to pick all the members. We use 2-body+SOAP descriptors, of cutoffs 6 \AA\ each, with 12 radial basis functions ($n=12$) and spherical harmonics expanded up to the 6th order ($l=6$) for SOAP. The descriptor scales ($\delta$) are set to 1 and 0.5 respectively. The energy and force sigma values (inverse weights) are the same in all cases: 0.001 eV/at. for energies and 0.03 eV/\AA\ for forces. The average predictions of the 15 models, along with the standard deviation, on the corresponding rigid molecule dissociations are presented with violet-to-yellow lines in panels \textbf{a2}, \textbf{b2} of Fig. \ref{fig:S6andS8diss-spinCommitteeEvolution-Rigid.png}. We observe a sharp increase in accuracy of ground state prediction going from 1 to 5 committee members (CM) and a plateau on reaching 15. 

Additionally, we see that the "Best of 1" mean curve predicts the wrong isolated dimers energies in both S$_{6}$ and S$_{8}$ cases. Train and test energy and force prediction errors for rattled molecule dissociations datasets composed of the best-of-$n$ calculations are shown in Fig. \ref{fig:S6andS8diss-spinCommitteeEvolution-Rigid.png} \textbf{c}, including for models trained on the combined S$_6$ and S$_8$ dataset. There is a clear decreasing trend for larger numbers of committee members, with the error values plateauing after a committee of size 5.

 To conclude, we showed that for two sulfur toy models of S$_6$ and S$_8$ molecules to dimers dissociations, where crucially the ground state of each structure is known, one can approach it by selecting the lowest energy of a committee of 15 DFT calculations with random initial atomic spins. This has implications for other spin-polarized systems, where optimal spin arrangement may not be known a priori. In addition, we give indication that a few points which are not at the minimum spin state (S$_8$ case in Fig. \ref{fig:S6andS8diss-spinCommitteeEvolution-Rigid.png}) do not break the fit. We reiterate that our approach models solely the ground state and would work best in cases where excited spin states are sparsely populated (e.g. away from the conical intersections of different spin surfaces at RC of 0.9 \AA\ in the S$_8$ rigid molecule dissociation in Fig. \ref{fig:S6andS8diss-spinCommitteeEvolution-Rigid.png} \textbf{b1} and \textbf{b2}). In comparison with spin-aware ML models, we do not have to feed in information about the atomic spins, and can use standard spin-agnostic ML fitting architectures. Regardless of the regression method, compared to non-spin-polarized cases, treatment of spin-polarized systems necessarily incurs a higher computational cost (aside from the factor of 2 due to spin-polarized DFT evaluation). In our case the multiplier is 15, as one has to evaluate a structure 15 times and pick the lowest energy result.
 
We are grateful for computational support from the UK national high performance computing service, ARCHER/ARCHER2, for which access was obtained via the UKCP consortium and funded by EPSRC grants ref EP/P022561/1 and EP/X035891/1. V.C. acknowledges support from the EPSRC Doctoral Training Programme grants EP/N509620/1 and EP/R513180/1.
The data that support the findings of this study are openly available in Apollo at \url{https://doi.org/10.17863/CAM.112212}\cite{dataset_doi}. G.C. has equity interest in Symmetric Group LLP that licenses force fields commercially, and also in Ångström AI.

\textbf{Vlad Cărare}: investigation, writing – original draft, methodology. \textbf{Volker Deringer}: conceptualization (equal), supervision (equal), writing - review and editing (equal). \textbf{Gábor Csányi}: conceptualization (equal), supervision (equal), writing - review and editing (equal).

\nocite{*}
\bibliography{aipsamp}

\end{document}